\documentclass{article} 
\usepackage{nips15submit_e,times}
\usepackage{hyperref}
\usepackage{url}
\usepackage{enumerate}

\title{Scalable Approximations of Marginal Posteriors in Variable Selection}

\author{
Willem van den Boom \\
Dept.\ of Statistical Science\\
Duke University\\
\texttt{willem.van.den.boom@duke.edu} \\
\And
Galen Reeves \\
Dept.\ of Electrical \& Computer Engineering and \\
Dept.\ of Statistical Science \\
Duke University\\
\texttt{galen.reeves@duke.edu} \\
\And
David B.\ Dunson \\
Dept.\ of Statistical Science \\
Duke University \\
\texttt{dunson@duke.edu} \\
}

\nipsfinalcopy 

\usepackage{amsmath}
\usepackage{amssymb}
\usepackage{amsthm} 

\usepackage{bm}
\usepackage[numbers]{natbib}
\usepackage{algorithm}
\usepackage[noend]{algpseudocode}
\usepackage{color} 
\usepackage{subcaption} 
\usepackage{graphicx} 
\usepackage{bbm} 

\begin{document}

\maketitle

\begin{abstract}
In many contexts, there is interest in selecting the most important variables from a very large collection, commonly referred to as support recovery or variable, feature or subset selection.  There is an enormous literature proposing a rich variety of algorithms.  In scientific applications, it is of crucial importance to quantify uncertainty in variable selection, providing measures of statistical significance for each variable.  The overwhelming majority of algorithms fail to produce such measures.  This has led to a focus in the scientific literature on independent screening methods, which examine each variable in isolation, obtaining $p$-values measuring the significance of marginal associations.  Bayesian methods provide an alternative, with marginal inclusion probabilities used in place of $p$-values.  Bayesian variable selection has  advantages, but is impractical computationally beyond small problems.  In this article, we show that approximate message passing (AMP) and Bayesian compressed regression (BCR) can be used to rapidly obtain accurate approximations to marginal inclusion probabilities in high-dimensional variable selection.  Theoretical support is provided, simulation studies are conducted to assess performance, and the method is applied to a study relating brain networks to creative reasoning.

\end{abstract}

\section{Introduction}

In many contexts, there is interest in selecting important variables from a very large collection. Think for instance of gene expression data or neuroimaging data, where the number of potential features (predictors) exceeds the sample size. To deal with this problem, many variable selection methods have been developed that scale well to large numbers of variables, with Lasso/$L_1$-penalization providing one example. A major downside of many of these fast methods is that they do not provide a well-defined measure of statistical significance. Being able to evaluate statistical evidence about whether a variable should be included or not is however crucial in making informed variable selection decisions.  In fact, in many biomedical applications, the main emphasis is on identifying which variables should be included, while reporting the level of evidence in the data that these are important variables; prediction is not directly of interest.

Modeling the data in a Bayesian fashion provides a natural framework to evaluate statistical evidence via the posterior. Even though many Bayesian variable selection methods exist~\cite{Mitchell1988}, they typically rely on Monte Carlo sampling for inference~\cite{George1993,George1997,Ohara2009}, which does not scale well with the number of candidate predictors. This led us to develop a general approximation framework for marginal posteriors in Bayesian linear regression, which can handle large numbers of candidate predictors by treating all but one of their coefficients as nuisance parameters and integrating them out (approximately) from the likelihood~\cite{Berger1999}. As an example, we will focus on the spike-and-slab prior for variable selection. Our method provides an estimate of the posterior probability of inclusion for each potential predictor.

Our contributions can be summarized as follows:
\begin{itemize}
	\item
	This paper presents a novel framework for marginal posterior approximation which is scalable and can handle highly non-Gaussian prior and posterior distributions.
	\item
It is shown how two state-of-the-art methods --- BCR and AMP --- can be used within the marginal approximation framework by providing  an approximation of the posterior predictive distribution of rotated data.
	\item 
	The framework is applied to the problem of Bayesian variable section with a spike-and-slab prior on both simulated data and a real study relating brain networks to creative reasoning.
\end{itemize}

\subsection{Bayesian Linear Regression} \label{sec:bayesian_linear_regression}

Consider the standard linear regression model, 
\begin{equation} \label{lm}
	y = X\beta+\epsilon, \quad \epsilon\sim\mathcal{N}(0,\sigma^2 I_n),
\end{equation}
where $X$ is a fixed $n\times p$ matrix of features, $\beta$ is a $p\times 1$ vector of unknown coefficients, $y$ is an $n\times 1$ response vector, $\epsilon$ is an $n\times 1$ residual vector and $\sigma^2$ is the error variance. Assuming that $\beta$ is drawn according to a prior distribution $\pi$, the posterior distribution obeys
\begin{align}
p(\beta | y ) \propto \exp\left( - \tfrac{1}{2 \sigma^2} \| y - X \beta\|^2 \right) \pi(\beta).
\end{align}
In practice, the joint posterior is difficult to visualize and interpret, and one routinely bases inferences on summaries of marginal posterior distributions for univariate functionals of the parameters.  The {\em marginal posterior distribution} of coefficient $\beta_j$ is obtained by marginalizing out the other coefficients $\beta_{(-j)}$, and is given by
\begin{align}
p(\beta_j  | y)  =  \int_{\mathbb{R}^{p-1}}  p(\beta | y )  \prod_{j' \ne j} d \beta_{j'} .
\end{align}
The {\em posterior predictive distribution} is the distribution of the response $y_\textnormal{new}$ to a new vector $x_\textnormal{new}$ conditional on having observed $(y,X)$, and is given by
\begin{align}
p(y_\textnormal{new}   | y)  \propto  \int_{\mathbb{R}^{p}} \exp\left( - \tfrac{1}{2 \sigma^2} ( y_\textnormal{new}  - x_\textnormal{new}^T \beta)^2 \right)  p(\beta | y )   d \beta .
\end{align}
As expressed above, both the posterior marginal distribution and the posterior predictive distribution require computing a high-dimensional integral over the posterior distribution. These integrals are challenging to compute in general, a massive literature has focused on  scalable approximation methods.

\subsection{Bayesian Variable Selection Using a Spike-and-Slab Prior} \label{sec:variable_selection}

The problem of variable selection is to identify which entries of the coefficient vector are nonzero. A standard approach in Bayesian variable selection is to assign a spike-and-slab prior on the coefficients having the form 
\begin{equation} \label{betaprior}
	\beta_j\overset{\text{iid}}{\sim} (1-\lambda)\delta_0+\lambda \, \mathcal{N}(0,\psi), \quad j=1,\dots,p,
\end{equation}
where $\delta_0$ is a point mass at zero, $\mathcal{N}(0,\psi)$ is a Gaussian distribution with mean zero and variance $\psi$, and $\lambda \in (0,1)$ is the prior inclusion probability. 

Let $\gamma$ be a binary $p \times 1$ vector where $\gamma_j$ is an indicator of the event $\{\beta_j \ne 0\}$. The posterior marginal inclusion probability of the $j$th coefficient is given by
\begin{align}
p(\gamma_j = 1)  = \frac{ \lambda \, p( y | \gamma_j = 1) }{ (1-\lambda) p(y |  \gamma_j = 0) + \lambda \, p(y |\gamma_j = 1) },
\end{align}
where $p(y| \gamma_j)$ can be expressed explicitly as:
\begin{align}
p(y| \gamma_j)  \propto 
  \sum_{ \gamma'  : \gamma'_j = \gamma_j }  \mathcal{N}(y |  0 , \Phi_{\gamma'}) \lambda^{k(\gamma')}   (1-\lambda)^{p-1 -k(\gamma')}, \qquad \Phi_\gamma = X_\gamma^T X_\gamma \psi + 
\sigma^2 I,
\end{align}
with $k(\gamma') = \sum_{j' \ne j} \gamma'_{j'}$.  Note that the marginal inclusion probabilities are available in a simple analytic form involving evaluation of normal densities and simple weights.  However, the complexity of the summation grows exponentially in $p$ and thus direct computation is infeasible for large $p$.

 MCMC and related sampling algorithms have been employed for posterior inference~\cite{George1993,George1997,Ohara2009}.  Such algorithms attempt to sample efficiently from the massive dimensional space of $2^p$ possible models; for $p$ even moderately large (in the 100s to 1000s) the space is so huge that there is no hope of visiting more than a vanishingly small proportion of the models. For example, $2^{100} \approx 1.27 \times 10^{30}$. This leads to high Monte Carlo error in estimating posterior model probabilities, with almost all models estimated to have zero probability (as they are never visited).  The topology of the model space makes efficient computation even more challenging, with local regions containing ``good'' models often separated by large regions containing relatively poor models.  Sampling algorithms have trouble efficiently moving between these isolated regions.  This has motivated a rich literature on better samplers~\cite{Nott2005,Clyde2011}.   However, one can argue that sampling is intrinsically intractable (e.g., even the best samplers can find a much better model after 10 million iterations), motivating our fast approximation approach.

\section{Posterior Marginal Approximation with IID Priors} \label{framework}

This section describes our framework for approximation of the posterior marginal distributions for an arbitrary iid prior on the coefficient vector $\beta$. 
\begin{enumerate}[1.]
\item  The first step is to apply a rotation to the observed data that decouples the dependence between an unknown coefficient of interest and the other unknown coefficients, which are viewed as nuisance parameters. This leads to a representation of the marginal posterior in terms of the posterior predictive distribution of a modified linear regression problem. 

\item The second step is to replace the posterior predictive distribution obtained in the first step with a (non-standard) Gaussian approximation. Interestingly, this approximation can be highly accurate even if the prior and posterior are highly non-Gaussian.  Recent techniques in the literature are used to compute the mean and variance. 
\end{enumerate}

\subsection{Connection Between Posterior Marginal and Posterior Prediction } \label{sec:connection}

This section describes how the posterior marginal distribution of the $j$th unknown coefficient can be expressed in terms of the posterior predictive distribution of a rotated regression problem.

For a fixed index $j$, consider the rotated data $(z,\tilde{y})$  defined by
\begin{align}
z = q_1^T y, \qquad  \tilde{y} = Q_2^T y,
\end{align}
where $q_1 = x_j / \| x_j\|$ is the unit vector in the direction of the $j$th column of $X$ and $Q_2$ is an $n \times (n-1)$ matrix chosen arbitrarily subject to the constraint that $Q_2 Q_2^T = I_n - q_1 q_1^T$. Since the $n \times n$ matrix $Q = [q_1 | Q_2]$ is full rank, the mapping between $y$ and $(z,\tilde{y})$ is one-to-one. To characterize these terms, we introduce the notation
\begin{align}
a = \|x_j\|, \qquad \tilde{x}_\textnormal{new} = q_1^T X_{(-j)}, \qquad \tilde{X} = Q_2^T X_{(-j)}, \qquad \tilde{y}_\textnormal{new}= z - a \beta_j. \notag
\end{align}
The first three terms are functions of $X$ and the last term is an auxiliary variable that cannot be observed directly, since $\beta_j$ is unknown. Following from the rotational invariance of the Gaussian distribution, the distribution of the rotated data can now be expressed as
\begin{align}
\tilde{y} | \beta &\sim \mathcal{N}\left( \tilde{X}\beta_{(-j)}    , \sigma^2 I_{n-1}\right), \label{eq:ytilde}  \\
\tilde{y}_\textnormal{new} | \beta  & \sim \mathcal{N} \left( \tilde{x}^T_\textnormal{new} \beta_{(-j)}  , \sigma^2 \right), \label{eq:ytildenew} \\
z & = a \beta_j + \tilde{y}_\textnormal{new}.
\end{align}
The important property of this decomposition is that $\tilde{y}$ does not depend on $\beta_j$. Thus, conditioned on $z$,  the posterior predictive distribution $p(\tilde{y}_\textnormal{new} |  \tilde{y})$ is a sufficient statistic for inference about $\beta_j$. 
This means that it is now sufficient to consider the scalar model 
\begin{align}
z = a \beta_j  + \tilde{y}_\textnormal{new}, \qquad \beta_j \sim \pi, \qquad \tilde{y}_\textnormal{new} \sim f,  \label{eq:scalar_model} 
\end{align}
where $f$ denotes the posterior predictive distribution $ p(\tilde{y}_\textnormal{new} |  \tilde{y})$. The posterior marginal distribution of $\beta_j$ obeys
\begin{align}
p(\beta_j | y) \propto f( z - a \beta_j) \pi( \beta_j). \label{eq:marginal_scalar} 
\end{align}

\subsection{Approximation of the Posterior Predictive Distribution} 

The main challenge in using the formulation of the previous section to efficiently approximate the marginal posterior of $\beta_j$ is that computation of the exact posterior predictive distribution is intractable.
The key insight underlying our approach is that, in many cases of interest, the posterior predictive distribution can be well-approximated by a Gaussian density, even if the prior and posterior distributions of the unknown coefficients are highly {\em non-Gaussian}. 

To obtain an approximation of $p(\beta_j | y)$, we propose to first obtain a Gaussian approximation $\hat{f}$ for the posterior predictive distribution $f$, and then plug this approximation into \eqref{eq:marginal_scalar}  to compute the approximation of the posterior on $\beta_j$,
\begin{align}
\hat{p}(\beta_j | y) \propto \hat{f}( z - a \beta_j) \pi( \beta_j).
\end{align}
The approximation $\hat{f}$ has the form $\mathcal{N}(\mu,\tau^2)$, where the mean $\mu$ and variance $\tau^2$ are functions of the data set $( \tilde{y}, \tilde{X}, \tilde{x}_\textnormal{new})$. The problem of computing $(\mu,\tau^2)$ can be attacked adapting a variety of recent techniques in the literature. Two of these are described in Sec.~\ref{sec:methods}.

\subsection{Approximation for Variable Selection} 

We now show how our marginal approximation framework can be applied to the problem of variable selection described in Sec.~\ref{sec:variable_selection}. Let $\pi$ be the spike-and-slab prior in \eqref{betaprior} and let $\mathcal{N}(\mu,\tau^2)$ be the approximation of $f$. The approximation for the posterior marginal distribution of $\beta_j$ is then a spike-and-slab distribution of the form
\begin{align}
 (1- \lambda_j ) \delta_0 + \lambda_j \, \mathcal{N}\left(m_j, \psi_j  \right), 
\end{align}
where $m_j = \frac{a \psi (z-\mu)}{a^2 \psi + \tau^2}$,  $\psi_j = \frac{ \psi \tau^2}{a^2 \psi + \tau^2}$, and  $\lambda_j$ is the approximation of the posterior marginal inclusion probability:
\begin{align}
\lambda_j & = \frac{ \lambda \, \mathcal{N}(z |\mu, a^2 \psi + \tau^2 ) }{  (1-\lambda) \mathcal{N}(z |\mu,  \tau^2 ) + \lambda \, \mathcal{N}(z |\mu, a^2 \psi + \tau^2 ) }.
\end{align}

\subsection{Analysis of Framework} 

A particularly  useful property of our approximation framework is that the discrepancy between the posterior predictive distribution $f$ and its approximation $\hat{f}$ does not depend on the unknown coefficient $\beta_j$ whose posterior marginal distribution we are trying to compute. As a consequence, if $\hat{f}$ converges to $f$ under a suitable metric, then it follows under very weak assumptions on the prior $\pi$ that the posterior marginal approximation also converges to the true posterior marginal.

As a heuristic justification for the Gaussian approximation of $\tilde{y}_\textnormal{new} | \tilde{y}$, consider a setting in which the entries of $\tilde{x}_\textnormal{new}$ are of the same order. Then, the a priori distribution of $\tilde{y}_\textnormal{new}$ is approximately Gaussian by the central limit theorem for sums of independent variables. Provided that the entries of $\beta_{(-j)}$ given $\tilde{y}$ are weakly correlated, it can then be argued that the posterior distribution of $\tilde{y}_\textnormal{new}$ is also approximately Gaussian. Using ideas from \cite{Bayati2011,bayati:2012}, this line of reasoning can be made rigorous for certain classes of large random matrices $X$. It is important to note that approximate Gaussianity of the predictive distribution does not hold in the setting where a small number of other predictors are highly collinear with $x_j$.

In contrast to many of the existing approximation methods, our framework can handle posteriors which are multimodal and posteriors which are discrete-continuous mixtures. This is not possible using methods based on direct normal-type approximations or Laplace's method~\cite{Kadane1986,Kass1990,Miyata2010}. 
Also, we note that previous work has shown how confidence intervals can be obtained for various M-estimators \cite{javanmard:2013}. Our work differs in that we can handle an arbitrary prior distribution, our two-stage procedure decouples the interaction between the coefficient of interest and the approximation, and our framework permits the use of a variety of methods to compute the posterior predictive distribution.

\section{Methods for Posterior Predictive Approximation}\label{sec:methods} 

The framework described in Sec.~\ref{framework} requires the approximation of the posterior predictive distribution of a rotated linear regression problem. This section shows how two recent methods  --- BCR and AMP --- can be used to obtain this approximation. Both of these methods are scalable and have theoretical performance guarantees in the high-dimensional setting. 

Throughout this section, the methods are described in the context of the usual posterior predictive distribution problem given in Sec.~\ref{sec:bayesian_linear_regression}. For the purposes of our approximation framework, however, it is important to remember that these methods are not applied to the original data $(y,X)$, but rather to the rotated data $(\tilde{y}, \tilde{X}, \tilde{x}_{new})$ defined in Sec.~\ref{sec:connection}.

\begin{algorithm}[tbp]
	\caption{Bayesian Compressed Regression (BCR) for posterior predictive distribution} \label{BCR}
	\textbf{Input:} data $(y,X)$, new vector $x_\textnormal{new}$, BCR parameters $(\kappa,m,K)$.
	
	\begin{enumerate}[1.]
		\item
		Run the BCR algorithm from \cite{Guhaniyogi2015} for $K$ random projections:
		\begin{algorithmic}[1]

			\For{$k=1,\dots,K$}
				\State Draw $\theta\sim\mathcal{U}(0.1,0.9)$
				\State Sample each element in the $p\times m$ matrix $\Theta$ from $(-\sqrt{1/\theta}, \sqrt{1/\theta}, 0)$ with probabilities $(\theta^2, (1-\theta)^2, 2(1-\theta)\theta)$ but such that $\Theta$ is full rank.
				\State Orthonormalize $\Theta$ using a Gram-Schmidt process.
				\State Compute the predictive mean and variance of new response $y_\textnormal{new}$ based on BCR with projection matrix $\Theta$ according to
				\begin{align*}
					\mu_k &= x_\textnormal{new} \Theta (\Theta^T X^T X \Theta + (\sigma^2/\kappa) I_m )^{-1} \Theta^T X^T y, \\
					\tau^{2}_k &= x_\textnormal{new} \Theta (\Theta^T X^T X \Theta + (\sigma^2/\kappa) I_m )^{-1} \Theta^T x_\textnormal{new}^T + \sigma^2.
				\end{align*}
				\State Compute the model weight $w_k$ which equals the marginal likelihood of the linear regression model $y\sim\mathcal{N}(X\Theta\alpha, \sigma^2 I_n)$ with prior $\alpha\sim\mathcal{N}(0,\kappa I_m)$
			\EndFor
		\end{algorithmic}
		
		\item
		Model average the means and variances of new response $y_\textnormal{new}$ according to
		\[
			\textstyle \mu = \sum_{k=1}^K w_k \mu_k, \qquad \tau^2 = \sum_{k=1}^K w_k \tau^2_k.
		\]
	\end{enumerate}
	
	\textbf{Return:} approximate posterior predictive distribution $\mathcal{N}(y_{new}| \mu,\tau^2)$.
	
\end{algorithm}

\subsection{Bayesian Compressed Regression}

BCR is inspired by data squashing and compressive sensing but is fundamentally different as it reduces only the number of predictors but not the sample size~\cite{Guhaniyogi2015}. It computes the exact posterior after randomly projecting the $n \times p$ matrix $X$ of features, using a random $p\times m$ ($m < p$) compression matrix $\Theta$, to the compressed $n \times m$ matrix with compressed features $X\Theta$. The coefficients $\beta$ are then replaced with an \mbox{$m$-dimensional} vector $\alpha$, which is assigned a normal prior.
We now have the likelihood $y\sim\mathcal{N}(X\Theta\alpha, \sigma^2 I_n)$.
Conditional on the projection, this then readily yields as posterior predictive for $y_\textnormal{new}|y$ a normal distribution $\hat{f}=\mathcal{N}(\mu,\tau^2)$. See Algorithm~\ref{BCR} for the details, using the prior $\alpha\sim\mathcal{N}(0,\kappa I_m)$ and the random projection from \cite{Guhaniyogi2015}.
As in \cite{Guhaniyogi2015}, we model average over $K$ different random projections.

For the spike-and-slab prior in \eqref{betaprior}, we will set $\kappa=\psi$ while $m$ can be chosen based on $\lambda$. To see the latter, the prior on $\alpha$ induces a singular prior on $\beta$ that lives in an $m$-dimensional hyperplane in $\mathbb{R}^p$ so a lower $m$ implies a more restricted $\beta$ which is a bit similar to using a smaller $\lambda$ in \eqref{betaprior}. Secs.~\ref{simulation} and \ref{application} use $K=10$.
\cite{Guhaniyogi2015} compressed all features to obtain a computationally efficient approach to prediction without the possibility of variable selection; here, we use BCR in a fundamentally new and different manner by leveraging the posterior approximation framework from Sec.~\ref{framework}.

The theoretical support for BCR applies to the high-dimensional setting where both $n$ and $p$ tend to infinity with $p\leq\exp(\delta n^\zeta )$ for $\delta>0$ and $\zeta\in(0,1)$. In this setting, concentration of the posterior approximation $\hat{f}$ is shown in \cite{Guhaniyogi2015} under some regularity conditions on $X$ and $\beta$. Combining this result with Theorem~1 in \cite{Jiang2007} shows that, under the spike-and-slab prior, the approximation $\hat{f}$ converges in probability to the true predictive distribution $f$.

\subsection{Approximate Message Passing}

The second method we consider is approximate message passing (AMP) \cite{donoho:2009a, Bayati2011, rangan:2011}. This algorithm is based on Gaussian and quadratic approximations of loopy belief propagation in graphical models.  It can also be viewed as a forward-backward primal-dual method for minimizing an approximation to the large-system-limit Bethe free energy \cite{rangan2013}. The algorithm is defined by scalar denoising functions, and optimal performance is obtained when this function is matched to the prior distribution of the coefficients \cite{reeves:2012}. 

For certain classes of large random matrices $X$ the behavior of AMP can be characterized rigorously as a function of the prior distribution on the coefficients \cite{Bayati2011,bayati:2012}. Furthermore, in these settings, simulations show that the algorithm converges rapidly and is often much faster than other state-of-the-art optimization methods. For general matrices, however, analysis of AMP is more challenging. In practice, convergence of the algorithm may require dampening \cite{rangan2013} or serial updates \cite{manoel:2015}. 

The output of AMP can be viewed as an approximation to the posterior marginal distribution of the coefficients, and is thus directly related the framework described in this paper. A key difference, however, is that we do not run AMP on the entire data. Instead, we first use AMP to obtain an estimate of the posterior predictive distribution of $y_\textnormal{new}$ and then combine this with the scalar measurement $z$. This two-stage procedure has the advantage that the first stage is independent of the coefficient of interest.  A high-level overview of our implementation of AMP for approximation of the posterior predictive distribution is given in Algorithm 2.

The theoretical support for AMP applies to the high-dimensional setting where both $n$ and $p$ tend to infinity with $n/p \rightarrow \delta$ for a fixed ratio $\delta \in (0,\infty)$. Then, under the assumption that the entries of $X$ are iid zero-mean Gaussian variables, it follows from results in \cite{Bayati2011,reeves:2012} that the approximation $\hat{f}$ converges in probability to the true predictive distribution $f$, provided that $\delta > \delta_0$ where $\delta_0 \in (0,1)$ depends on the prior $\pi$ and the error level $\sigma^2$.

\begin{algorithm}[btp] \label{alg:AMP} 
	\caption{Approximate Message Passing (AMP) for posterior predictive distribution} \label{AMP}
	\textbf{Input:} data $(y,X)$, new vector $x_\textnormal{new}$, prior distribution $\pi$, error variance $\sigma^2$. 
	\begin{enumerate}[1.] 
	\item Run sum-product approximate message passing with optimal nonlinearity defined by the prior $\pi$ and obtain point estimates of the posterior mean $m$ and posterior marginal variance $v$ of the regression coefficients. For full details, see e.g.~\cite[Algorithm 1]{rangan2013} or \cite[Sec. 2.5 ]{parker:2014}.
	\item Compute the mean and variance of new response $y_\textnormal{new}$ according to
	\begin{align}
	\mu = x_\textnormal{new}^T m, \qquad \tau^2 =  x_\textnormal{new}^T \operatorname{diag}(v) x_\textnormal{new} + \sigma^2. \notag
	\end{align}
	
	\end{enumerate}
	\textbf{Return:} approximate posterior predictive distribution $\mathcal{N}(y_\textnormal{new}| \mu,\tau^2)$.
\end{algorithm}

\section{Simulation Studies} \label{simulation}

\begin{figure}[tb]
	\centering
	\includegraphics[width=.5\textwidth]{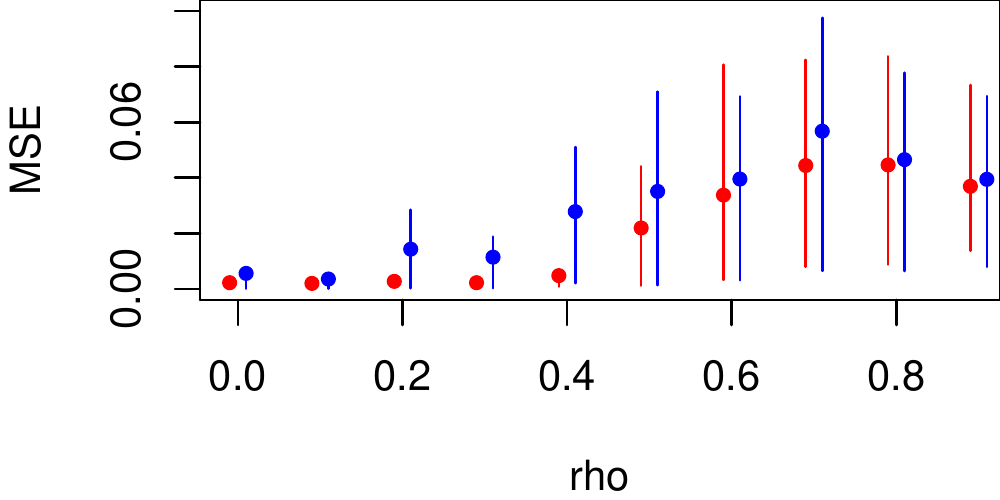}
	\caption{Average (dot) and 20th to 80th percentile (line) of the MSE of the posterior inclusion probabilities estimates by the AMP and BCR algorithm in red on the left and blue on the right, respectively.}
	\label{simul1}
\end{figure}

To test our methods to approximate posterior inclusion probabilities, we apply them to an example where we can compute the exact inclusion probabilities. Consider the linear model $y\sim\mathcal{N}(X\beta,\sigma^2 I)$ with $n=100$, the $p=12$ parameters $\beta = (3,1.5,2,0,0,0,0,0,0,0,0,0)^T$ and the matrix with features $X$ such that its columns are identically normally distributed with correlation $\rho^{|i-j|}$ between column $i$ and $j$ and the elements within a column are iid, similar to Example~1 from \cite{Tibshirani1996}. According to this, we generate 100 data sets for each $\rho=0,0.1,\dots,0.9$ with $\sigma^2$ such that the signal-to-noise ratio equals 2 and we set $\psi = 10\sigma^2$. Then we compute the exact posterior inclusion probabilities and run our algorithm with AMP and with BCR ($m=5$) where $\lambda = 3/12$. Summaries of the MSE of the inclusion probabilities provided by our algorithms compared to the true posterior inclusion probabilities are shown in Fig.~\ref{simul1}.
We see that both approximation methods do a good job for $\rho=0$. As the collinearity in $X$ increases, the MSE increases as well but is still sufficiently small for the inclusion probability estimates to be meaningful.

In practice, $\sigma^2$ and $\lambda$ are parameters that need to be tuned. For the application in the next section, we build this tuning into our algorithms. For AMP, we use iterative updating of $\sigma^2$ very similar to what is proposed in \cite{Vila2011} and also update $\lambda$ analogously based on the current inclusion probability estimates at each iteration. BCR allows for marginalizing out $\sigma^2$~\cite{Guhaniyogi2015}. For this, we use a $\sigma^2\sim\mathcal{IG}(3,1)$ prior after standardization of both $X$ and $y$. We then iterate the BCR algorithm where each time we update $\lambda$ in the same manner as for the AMP algorithm. Convergence of $\lambda$ required only a few iterations.

\begin{figure}[htbp]
	\centering
	\begin{subfigure}[t]{.17\textwidth}
		\centering
		\includegraphics[width=.99\textwidth]{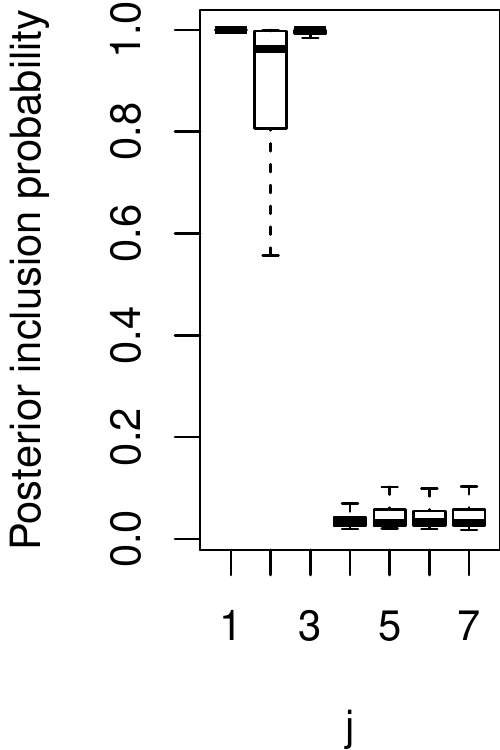}
		\caption{ \centering $\rho = 0$, $\mathrm{SNR}= 1$}
	\end{subfigure}
	~
	\begin{subfigure}[t]{.17\textwidth}
		\centering
		\includegraphics[width=.99\textwidth]{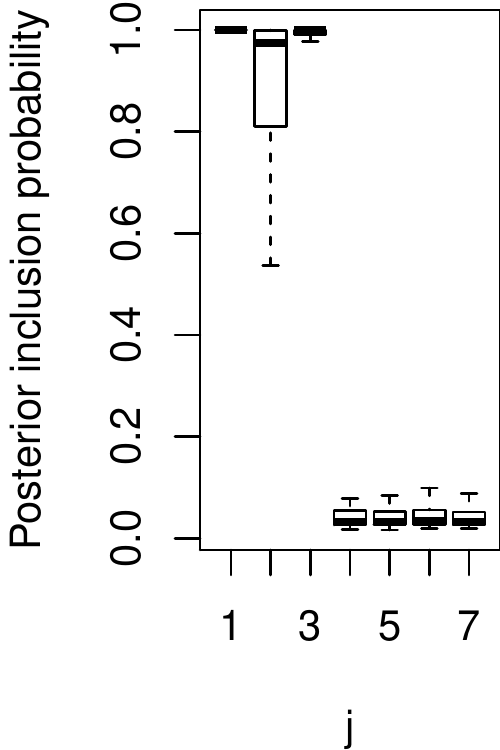}
		\caption{\centering $\rho = 0.2$, $\mathrm{SNR}= 1$}
	\end{subfigure}
	~
	\begin{subfigure}[t]{.17\textwidth}
		\centering
		\includegraphics[width=.99\textwidth]{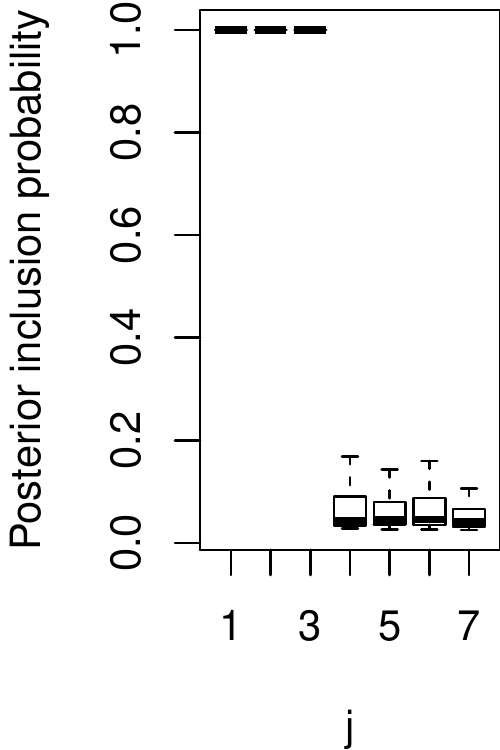}
		\caption{\centering $\rho = 0.5$, $\mathrm{SNR}= 10$}
	\end{subfigure}
	~
	\begin{subfigure}[t]{.17\textwidth}
		\centering
		\includegraphics[width=.99\textwidth]{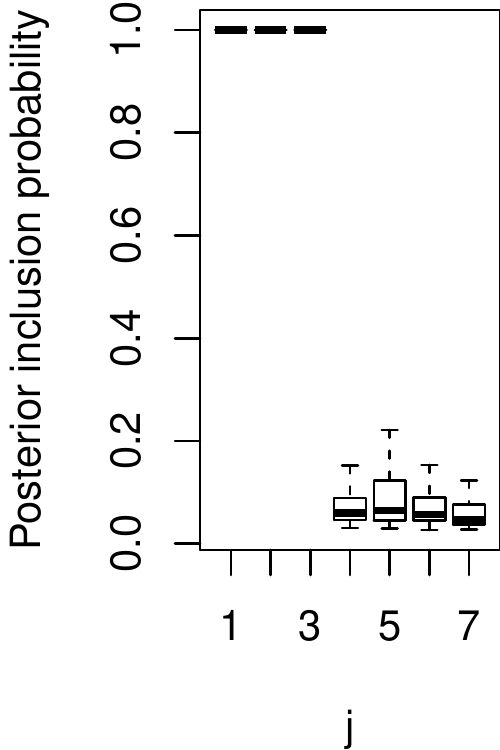}
		\caption{\centering $\rho = 0.7$, $\mathrm{SNR}= 10$}
	\end{subfigure}
	~
	\begin{subfigure}[t]{.17\textwidth}
		\centering
		\includegraphics[width=.99\textwidth]{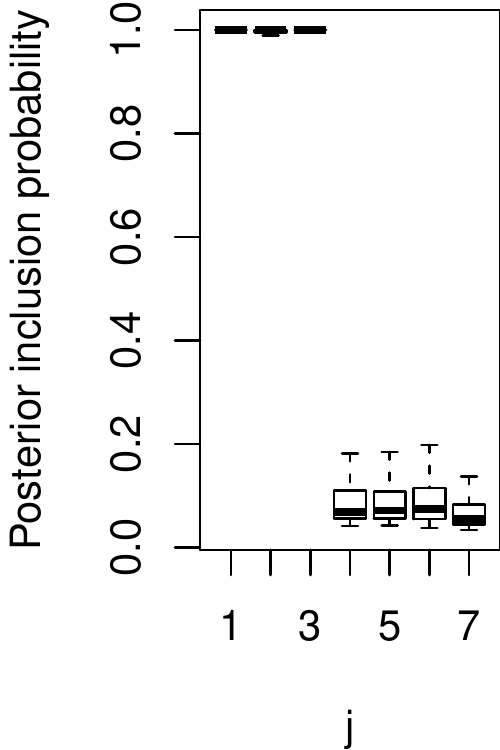}
		\caption{\centering $\rho = 0.8$, $\mathrm{SNR}= 10$}
	\end{subfigure}
	\caption{Results for BCR with $\sigma^2$ and $\lambda$ unknown and with $j$ the index of the parameter.}
	\label{simul2BCR}
\end{figure}

\begin{figure}[htbp]
	\centering
	\begin{subfigure}[t]{.17\textwidth}
		\centering
		\includegraphics[width=.99\textwidth]{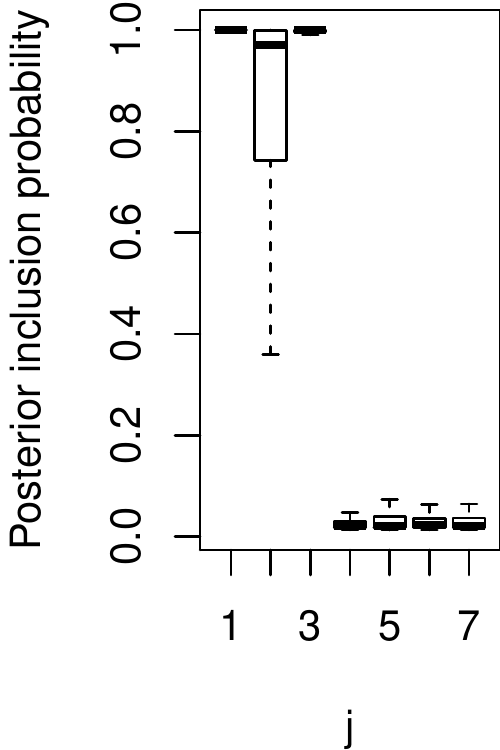}
		\caption{\centering $\rho = 0$, $\mathrm{SNR}= 1$}
	\end{subfigure}
	~
	\begin{subfigure}[t]{.17\textwidth}
		\centering
		\includegraphics[width=.99\textwidth]{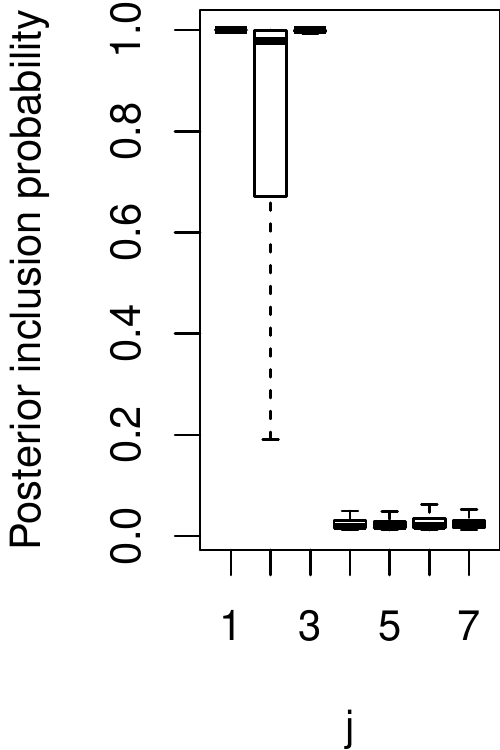}
		\caption{\centering $\rho = 0.2$, $\mathrm{SNR}= 1$}
	\end{subfigure}
	~
	\begin{subfigure}[t]{.17\textwidth}
		\centering
		\includegraphics[width=.99\textwidth]{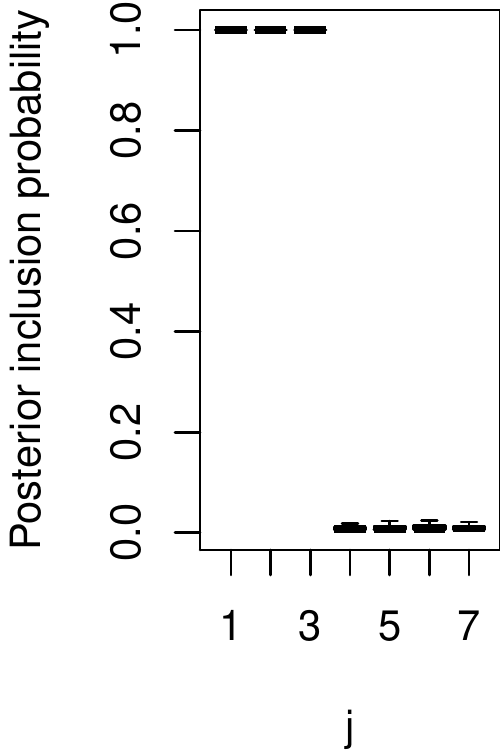}
		\caption{\centering $\rho = 0.5$, $\mathrm{SNR}= 10$}
	\end{subfigure}
	~
	\begin{subfigure}[t]{.17\textwidth}
		\centering
		\includegraphics[width=.99\textwidth]{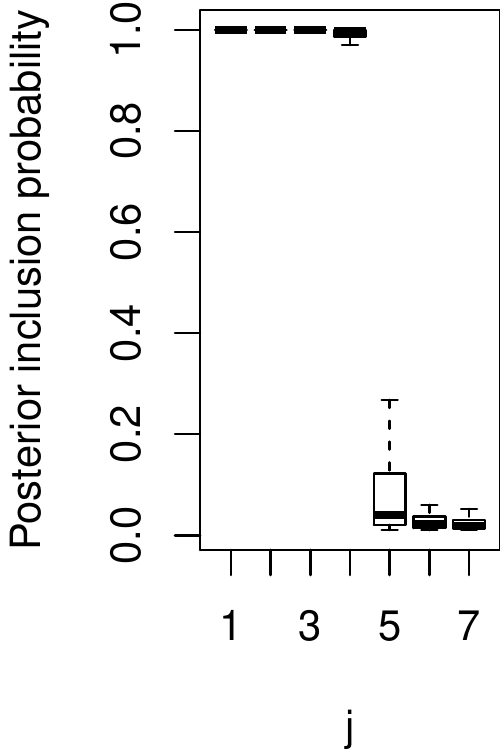}
		\caption{\centering $\rho = 0.7$, $\mathrm{SNR}= 10$}
	\end{subfigure}
	~
	\begin{subfigure}[t]{.17\textwidth}
		\centering
		\includegraphics[width=.99\textwidth]{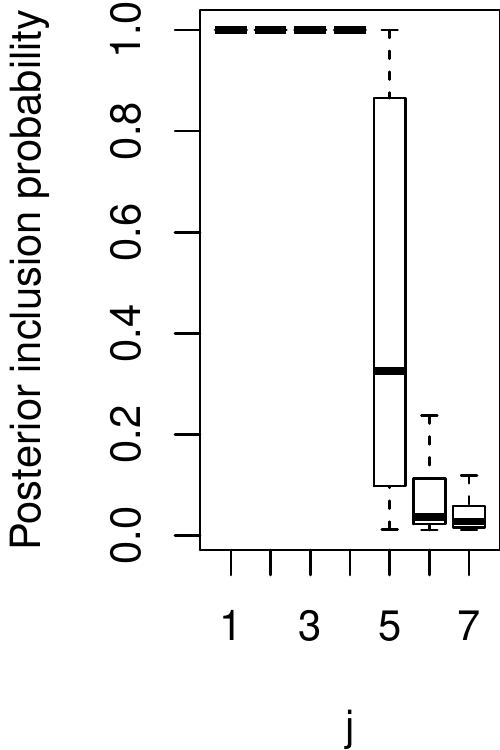}
		\caption{\centering $\rho = 0.8$, $\mathrm{SNR}= 10$}
	\end{subfigure}
	\caption{Results for AMP with $\sigma^2$ and $\lambda$ unknown and with $j$ the index of the parameter.}
	\label{simul2AMP}
\end{figure}

To check whether these additions to our methods are valid, we created the box plots in Figs.~\ref{simul2BCR} and \ref{simul2AMP}, each based on 200 simulated data sets.
The design matrix is generated as earlier in this section but now $p = 7$ with $\beta = (3,1.5,2,0,0,0,0)^T$. Note that AMP is more aggressive than BCR in setting inclusion probabilities to the extreme values 0 and 1. Furthermore, AMP struggles with $\rho=0.7,0.8$. The results show that our methods can provide meaningful posterior inclusion probabilities even when $\lambda$ and $\sigma^2$ are unknown. 

\section{Applications} \label{application}

This section applies our approximation framework to neuroscience data. We use the brain network data from \cite{Jung2009} which is available as the MRN-111 data set from \url{http://openconnecto.me/data/public/MR/MIGRAINE_v1_0/}. These are connectomes with counts of the number of connections between the 70 different brain regions from the Desikan atlas~\cite{Desikan2006}. The dependent variable is the composite creativity index (CCI) for $n=113$ persons from \cite{Jung2009}. Out of all pairs of brain regions, 1802 have a connection for at least on person. The connection counts for these $p=1802$ pairs form the linear predictors in our model. These counts are zero for half of these pairs with a mean of 1477 and a maximum of 58090.

\begin{figure}[tb]
	\centering
	\begin{subfigure}[t]{.4\textwidth}
		\centering
		\includegraphics[width=.99\textwidth]{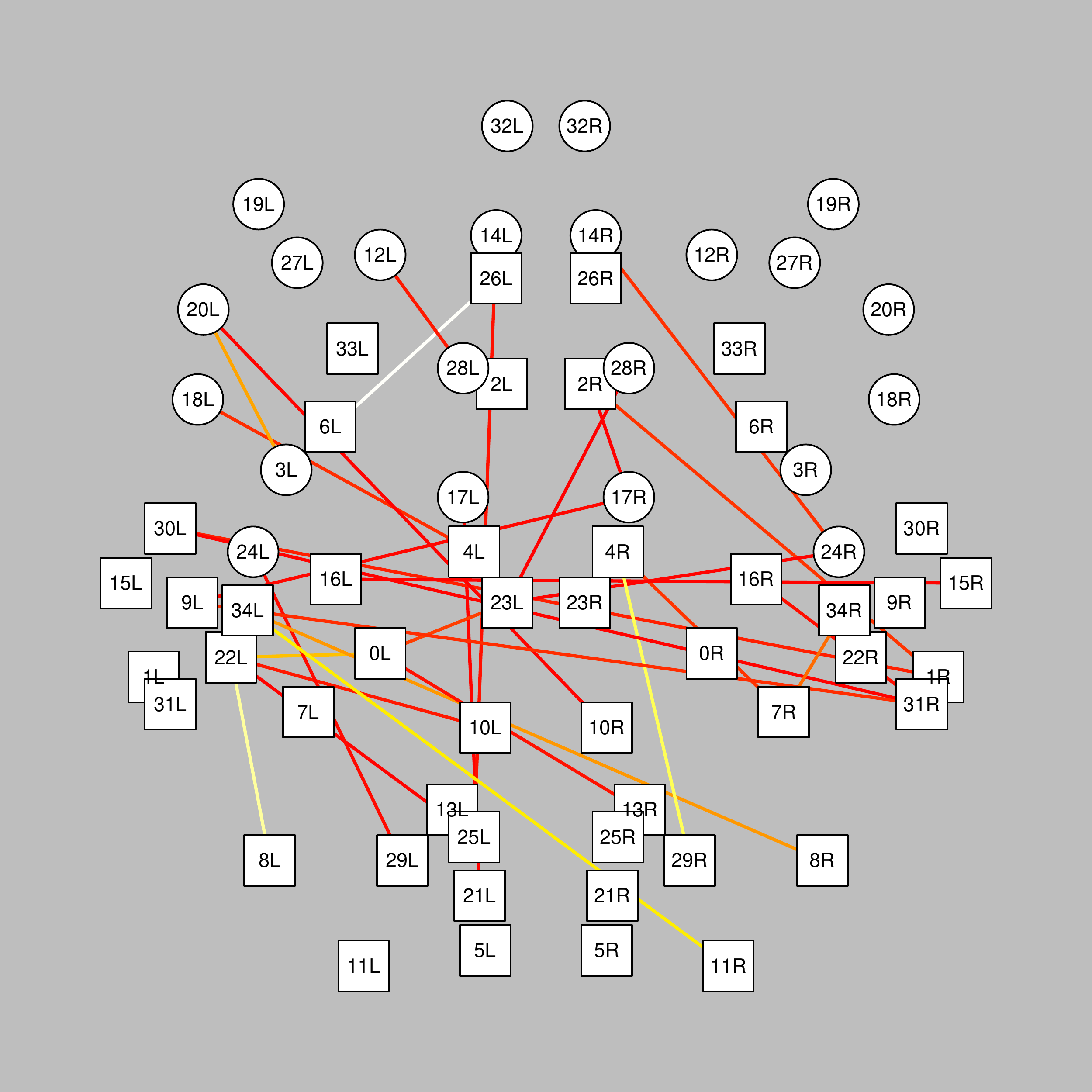}
		\caption{BCR}
	\end{subfigure}
	~
	\begin{subfigure}[t]{.4\textwidth}
		\centering
		\includegraphics[width=.99\textwidth]{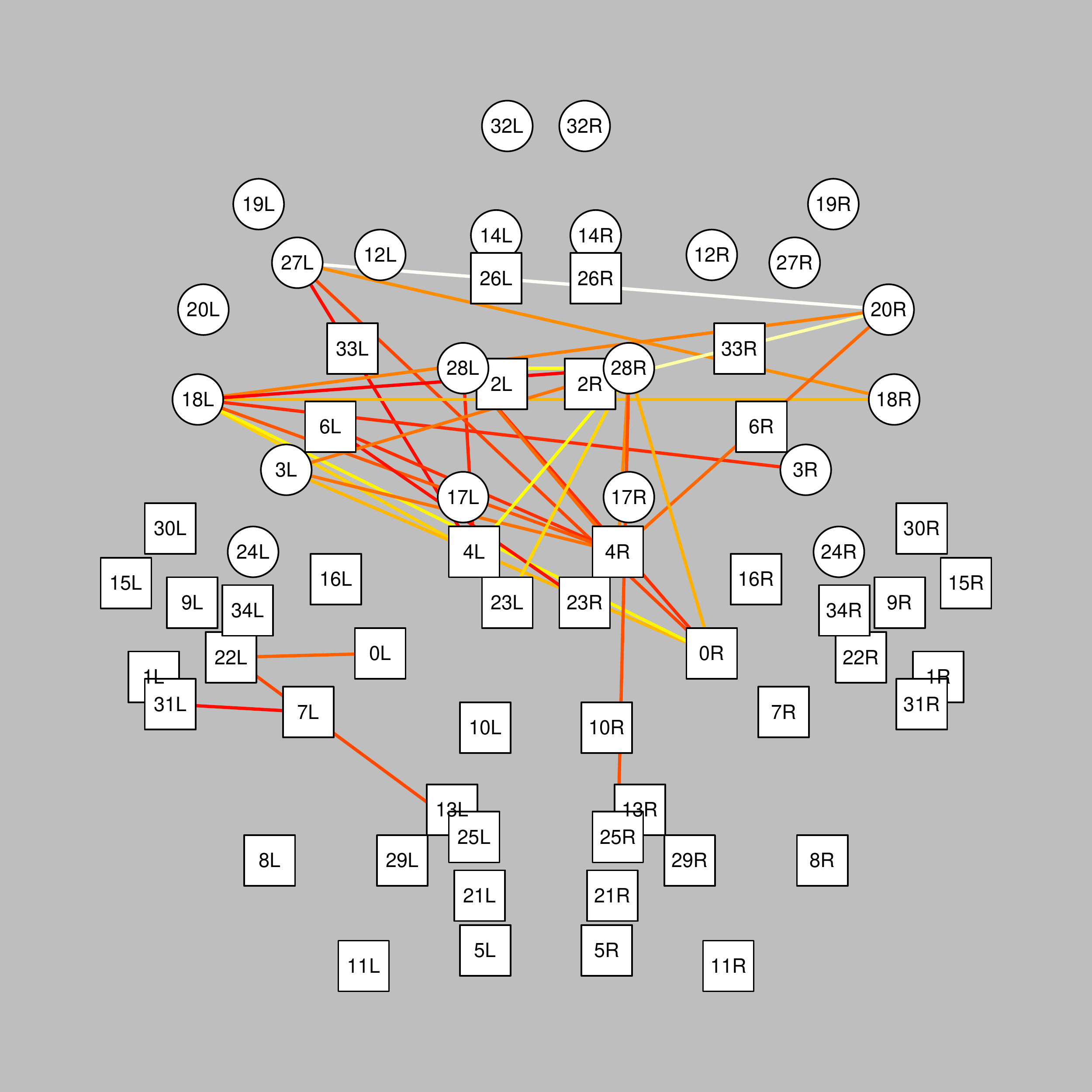}
		\caption{AMP}
		\label{resultsAMP}
	\end{subfigure}
	\caption{Horizontal section of a weighted brain network visualization where the weights are given by the posterior inclusion probability of each edge with red to white indicating low to high. For each algorithm, only the 60 edges with the highest inclusion probabilities are plotted.}
	\label{results}
\end{figure}

See Fig.~\ref{results} for a summary of the inclusion probability estimates provided by AMP and BCR ($m=20$). It is noteworthy that Fig.~\ref{resultsAMP} has similarities with Fig.~10 from \cite{Durante2014} which considers the same data set but using a Bayesian nonparametric model that is substantially more complex. For instance, both contain edges from node 18L to 3R, 18R and 20R. We also note that this is a highly challenging example due to the ill-conditioned, discrete and heavy-tailed nature of the feature matrix.  This likely leads to some of the differences between the BCR and AMP-based approaches, though both tend to include more cross-hemisphere connections, which is consistent with previous evidence that more creative individuals have more connections between the right and left hemispheres.

\section{Discussion}

Sec.~\ref{framework} presented a novel general framework for marginal posterior approximation. Via a rotation, the parameter of interest and the other `nuisance' parameters are separated in the likelihood. This reduces the $p$-dimensional problem to a scalar one dependent on the influence of the nuisance parameters. This influence, summarized in a posterior predictive, appears to be well approximated by a Gaussian for large $p$ even when the full posterior is far from Gaussian. Sec.~\ref{sec:methods} provided BCR and AMP as state-of-the-art methods for approximating this posterior predictive. We then focused on the spike-and-slab prior but note that the framework readily applies to many iid priors on $\beta$. The first simulation in Sec.~\ref{simulation} showed that the framework both with BCR and AMP is able to estimate the posterior inclusion probabilities accurately.

The proposed approach represents a substantial paradigm-shift in methods for estimating marginal posterior distributions in variable selection.  There is an enormous literature proposing a wide variety of carefully designed sampling algorithms; the proposed novel framework leads to order of magnitude speed ups and improvements in stability.  This same approach should be applicable much more widely than the Gaussian linear regression setting considered here.

\subsubsection*{Acknowledgments}

This material is based upon work supported in part with funding from the Laboratory for Analytic Sciences (LAS). Any opinions, findings, conclusions, or recommendations expressed in this material are those of the authors and do not necessarily reflect the views of the LAS and/or any agency or entity of the United States Government.
The authors would like to thank Rex E. Jung and Sephira G. Ryman for the brain connectivity data and creativity scores funded by the John Templeton Foundation (Grant 22156) entitled ``The Neuroscience of Scientific Creativity.''

\subsubsection*{References}

\begingroup 
\renewcommand{\section}[2]{}

\small{
\bibliographystyle{cj}
\bibliography{nips2015.bib}
}

\endgroup

\end{document}